# Semantic Degrees for Industrie 4.0

Deciding on the degree of semantic formalization to select appropriate technologies

Chih-Hong Cheng, Tuncay Guelfirat, Christian Messinger, Johannes Schmitt, Matthias Schnelte, Peter Weber [*]

ABB Corporate Research
Wallstadter Straße 59
68526 Ladenburg, Germany
{firstname.lastname}@de.abb.com

*Abstract* Under the context of Industrie 4.0 (I4.0), future production systems provide balanced operations between manufacturing flexibility and efficiency, realized in an autonomous, horizontal, and decentralized item-level production control framework. Structured interoperability via precise formulations on an appropriate degree is crucial to achieve engineering efficiency in the system life cycle. However, selecting the degree of formalization can be challenging, as it crucially depends on the desired common understanding (semantic degree) between multiple parties. In this paper, we categorize different semantic degrees and map a set of technologies in industrial automation to their associated degrees. Furthermore, we created guidelines to assist engineers selecting appropriate semantic degrees in their design. We applied these guidelines on publically available scenarios to examine the validity of the approach, and identified semantic elements over internally developed use cases targeting semantically-enabled plug-and-produce.

## I. Introduction

Industrial manufacturing companies are facing strong demands to improve their production process, not only on the shop-floor but throughout the complete value chain. These demands arise from requests such as growing productivity expectations, increasing number of product variants, reducing lot sizes, etc. It is widely perceived that new information technologies will reshape production processes via an integration into existing industrial automation and communication technologies, from engineering to commissioning to operation. The activity of Industrie 4.0 (I4.0, or Industrial Internet of Things) is such an initiative to apply internet of things (IoT) technologies to the industrial manufacturing context.

Nevertheless, to enable automated interpretation and processing of the interchanged information throughout the enterprise (from machines to services), apart from basic physical communication and data transmission, having a *commonly exposed information for data exchange* is a premise. We refer the degree of common understanding as the *semantic degree*. A predefined semantic-degree governs engineering and commissioning (e.g., implementation or configuration) efforts, as it is highly associated with the amount of information to be revealed on the functional interface.

Therefore, to agree on a proper semantic degree is crucial. The risk of improper "selection of semantic degrees" can be observed by two extremes: On one side, choosing a degree of no common understanding implies that each component needs to implement its own parser to understand the meaning of other components. On the other extreme, having a degree where each component understands the real world (in terms of physical equations) simply contains unnecessary details and incurs huge reasoning effort.

In this paper, we categorize the exposed semantic degree (Section II) by first considering whether the exchanged information is structural or behavioral (i.e., the receiver needs to know the system configuration which evolves over time). For structural information, semantic degrees range from simple document repositories to ontologies, while for dynamic information, finite automata, Petri nets, or genetic programming language are used. We also provide a categorization how commonly seen technologies in industrial automation are mapped into their associated degrees (Section III). With predefined degrees, we further identify a set of guidelines that can be used as a filtering procedure, enabling engineers to select appropriate semantic degrees in their system design (Section IV).

Based on these guidelines, we examine publically available "I4.0 demonstrators" and identify the required semantic degrees to fulfill the implemented features (Section V). One surprising result is that nearly all demonstrated features in our investigated demonstrators do not need to employ complex semantic degrees such as ontologies, i.e., it is largely sufficient to use glossary (controlled vocabulary) in the implementation.

To demonstrate the real usage of semantics, we lastly detail four internally-developed use cases from different industrial segments and transform them into future I4.0 scenarios (Section VI). These scenarios characterize intelligent machines or devices that support a "semantically-enabled plug-and-produce" metaphor to enable flexible production, simplified engineering and easy reconfiguration. We identify the ingredients and detail the required semantic information modeling, demonstrating its cruciality for engineering efficiency in the product life cycle, from design to operation to maintenance to renewal. We summarize related work and conclude in Section VII and VIII.

---

[*] Author list in alphabetical order

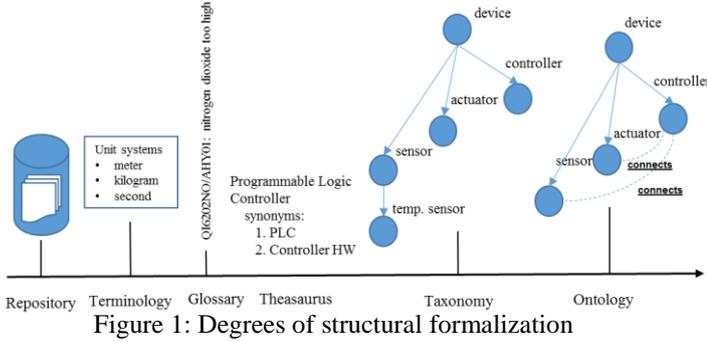

Figure 1: Degrees of structural formalization

## II. CATEGORIZATION CONCEPT

In order to categorize semantic degrees, we first distinguish between structural and behavior modelling. While structural modelling describes **how** (in which detail) information is exposed, behavioral modelling describes **what** kind of information is contained and knowledge about the information processing. These two aspects can be seen as two dimensions with certain degrees.

Based on the demands of transparency and collaboration that one wants to achieve in the system, these degrees help engineers to select an appropriate modelling method and technology.

### A. Structural formalization

Our categorization of structural information (Figure 1) is based on a refined set of criteria from the work of Navigli and Velardi [13], ranging from document repository to ontology. Table I provides their exposed information and some applied industrial contexts. Notice that our definition of the degree elements (e.g., glossary) are slightly altered to fit into the automation setup.

TABLE I. FEATURES PROVIDED BY STRUCTURAL DEGREE

| Structural Degree | Exposed Information | Industrial Context |
|---|---|---|
| S0: Repository | Unstructured/structured data | Devices catalogues, file structure, |
| S1: Terminology | Controlled vocabulary | Tags, annotations |
| S2: Glossary | Description over vocabularies | Human readable documentation, HMI |
| S3: Thesaurus | Basic relationships (association), similarities | Profile mapping, technology Integration |
| S4: Taxonomy | Tree Structure, parent-child relations, classifications | Abstraction (typing), device classification |
| S5: Ontology | Typed elements and relations | Interfaces, inheritance, topologies |

Starting from the lowest degree of formalization, a **Repository** can be seen as a source for raw data (either structured or unstructured) without additional semantics. This means that additional knowledge is needed for interpreting the information. A **Terminology** provides a set of vocabularies that is implicitly controlled, agreed and specialized for the system under investigation – which means that a term is well defined and unique. This makes it possible to annotate elements with tags in order to make information out of data. A data point for example can hold the value: [33.3] and provide an additional tag: [Celsius] – a controlled application of this tag allows for comparison with other values of the same kind. A **Glossary** is an explicit, specialized list of words and their associated definitions. Within industrial contexts, it can be used for extending elements (tags) with a human readable description, such as associate an alarm tag with its actual meaning that can be shown in an HMI (e.g., associate tag QI6202NO/AHY01 with meaning "nitrogen dioxide too high„). Importantly, we assume that the use of basic words such as measurement units (e.g., Kilogram, Meter), are considered as a common knowledge in terminology. However, these words do not need to be listed in a glossary, because of their roots in the common-sense.

A **Thesaurus** (or a Topic Map) is used to describe similarities between tags and to assign a single, well-defined term to all occurrences (e.g. from [Temperature Sensor] to [Bluetooth-Device_000A3A58F310]). It can also be used to map different technology terms from different vendors to avoid misunderstandings. **Taxonomies** are used to define parent-child relations and to build up tree structures, which are used in structuring device classes (e.g. a dimmer is a switch, a switch is an actuator). In contrast to taxonomies, an **Ontology** can have a full mashed topology. An ontology provides typed elements, references and enables the definition of type structures – which can be applied for the definition of interfaces as known from object models. Objects or object models (e.g. as used for the Common Information Model – CIM [31]) can be also represented by an ontology (however an object model typically has a fixed set of reference types).

### B. Behavioral formalization

The entries in TABLE II. describe behavioral aspects which may be required to be exposed by a certain scenario in the industrial context[2]. Contents from B0 to B2 are more static – they are basic ingredients used to record the snapshot of the system and to specify system invariants (conditions where the system should hold in certain stages), while contents from B3 to B6 target to expose dynamic information, i.e., they are used when understanding how system evolves over time is needed.

TABLE II. FEATURES PROVIDED BY BEHAVIORAL DEGREE

| Behavioral Degree | Exposed Information | Industrial Context |
|---|---|---|
| B0: Data | Values | Data Points |
| B1: Information | Tags | Self description, Data types |
| B2: Constraints (predicates over information) | Requirements, properties | Structural, logical policies, conditions to trigger alarms, simple logic (if-then-else rules) |

---

[2] The selection of elements in the behavioral degree is based on whether one can easily find appropriate industrial contexts. E.g., one can also insert pushdown automata in the behavioral degree, but as pushdown systems are rarely used, it is thus not listed.

| B3: Finite automata | States, events | State machines |
| --- | --- | --- |
| B4: Petri Nets | Signals, concurrencies | Concurrent/distributed processes, queues |
| B5: Programming Language | Flexible function set | Scripting, coding |
| B6: Integrated simulation model | Physical and logical behavior combined | Virtual commissioning, run-time optimization (e.g., MPC) |

The basic behavioral degrees are provisioning of **Data** and **Information**. While the first degree provides only values, the second extends the first with a meaning, which is needed to make information out of data. Data can also include so called BLOBS (binary large objects) - where also complex information can be contained but however without any (machine interpretable) relation to other parts of the model. **Constraints** can be used to formalize logical requirements over the information such that they be evaluated (to true or false) during commissioning or operation in finite amount of time[3]. E.g., ConNO > THREA is a predicate checking if the concentration of Nitro-Dioxide (ConNO) exceeds a pre-defined threshold value (THREA). In an industrial context, a system can use constraints to specify logical policies for their installation (e.g. size, energy consumption) or structural policies for their integration (e.g. allowed sub-modules or connectivity to other systems). By exposing constraints, one is able to perform automatic reasoning using state-of-the-art tools like SMT [4], SAT [1], or DL [12] reasoning solvers.

In some cases, exposing the information as a **Finite automaton (FA)** can become relevant – e.g. if a machine is supposed to expose its current state and potential state-changes to preceding or subsequent machines in order to support them to react accordingly. The state of a finite automaton, as presented in the Kripke structure, can actually be evaluations over atomic propositions where each atomic proposition is a constraint that can be algorithmically evaluated. The degree of **Petri Nets (PN)** becomes relevant if distributed or parallel processes have to be coordinated through the information model because of shared resources. The almost highest degree for the description of behavior is the use of a (general purpose) **Programming Language (PL)**, whose underlying model of computation can be viewed as Turing machines.

Still, for modeling and correctly interacting with physical environments, it is well known that models such as FA or PN lack the explicit notion of time [8]. An **Integrated Simulation Model (ISM)** combines the logic of a component like the programmable behavior and the surrounding physical environment. Examples include Modelica [7] or Ptolemy II [5]. With this degree, a system can derive the behavior of the physical process and use this knowledge in automatic decision support, such as the run-time optimization via model-predictive control (MPC) [2].

---

[3] Precisely, we only consider fragments of logic having decidability results. Thus, generic first-order logic (FOL) is considered as inappropriate, while decidable fragments of FOL such as description logic (DL [12]; used in OWL reasoning) is allowed.

### C. Integration of structural and behavioral modelling

Structural and behavioral modelling have to be integrated in order to allow for the combination of their features. For example *constraints* from the behavior model can make use of *tags* defined in the structural model – a climate control can restrict its input for temperature sensor values to [Celsius]. As another example a*utomata* in the behavior model can be linked with sensor values of a specific data type considering inheritance described by *ontologies* in the structural model.

### III. SELECTING THE APPROPRIATE DEGREE FOR STRUCTURAL FORMALIZATION

In our semantic degree formulation, usually a model can be mapped into a higher degree without the loss of information. The question now could be – why not simply take the highest degree? As shown in Figure 2, information modelling is a tradeoff between modelling and integration efforts. One would like to reduce the cost of one-time modeling efforts, while the result of modeling is still sufficient to allow efficient integration and commissioning, which is done per installment and can appear multiple times.

The decision process for the appropriate structural and behavior modelling starts with the analysis of the requirements of the application scenario. We provide some generic and exemplary guidelines in TABLE III, such that an engineer can use it to analyze its application scenario and derive the appropriate degree quickly. Notice that when all elements that are unique or rarely used, creating a detailed modeling for information exchange turns inefficient (see also Rule R0).

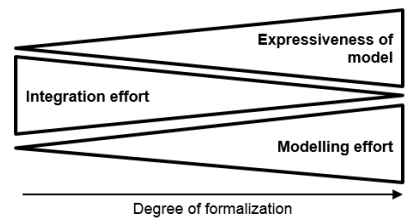

Figure 2: Tradeoff between modelling and integration efforts

TABLE III.  EXAMPLARY GUIDELINES FOR SELECTING THE APPROPRIATE DEGREES

| Rule | Minimal degree when the answer is positive | Argument |
| --- | --- | --- |
| R0: Is the scope of the system very limited? E.g. one vendor, only few entitities, static setup | S0: Repository B0: Data | Hard coding with less effort than modelling |
| R1: Have multiple parties the need to exchange standardized knowledge - which can be intuitevly understood (such as units) ? | S1: Terminology B1: Information | Necessity of well defined terms. |
| R2: Have multiple parties the need to coordinate the use of terms. | S2: Glossary B1: Information | Human readable description needed for a common understanding |
| R3: Is it necessary to integrate definition of terms of other parties | S3: Thesaurus B1: Information | Mapping of different definitions using a Thesaurus |

| R4: Should the system provide a basic type system and be extensible in terms of lately added types? | S4: Taxonomy B1: Information | Parent-Child relations needed to classify types. |
|---|---|---|
| R5: Should the system be dynamic and extensible – e.g. allow for modelling of artifical elements during runtime? | S5: Ontology B1: Information | Even the meaning of a relationship can be modelled |
| R6: Is it required to validate evolving configurations during runtime? | S1: Terminology B2: Constraints | Modelling of requirements and a controlled vocabulary needed |
| R7: Shall reasoning be supported? | S5: Ontology B2: Constraints | Necessity to describe complex and evaluable relationships |
| R8: Is it necessary to understand/modify the functionality (e.g. logic) of another system? | S1: Terminology B3:Automata (up to B5) | Machine interpretable description of logic required |

## IV. TECHNOLOGIES IN AUTOMATION AND THEIR UNDERLYING SEMANTIC DEGREES

For our proposed semantic degrees, we have examined existing (software-related) technologies in industrial automation and associate each technology with the corresponding degree. The result is shown in Table IV; it can be used by software engineers in industrial automation to quickly filter technologies to be used in their I4.0 projects. E.g., if CAEX is used during the communication of I4.0 entities, an ontology system is communicated underneath.

TABLE IV. TECHNOLOGIES IN AUTOMATION AND THEIR CORRESPONDING SEMANTIC DEGREES

| Automation Technologies | | Structural Degree | Behavioral Degree |
|---|---|---|---|
| GSD/GSDML/CFF/IODD/ESI/SCL | | S1: Terminology | B1: Information |
| MQTT | | S0: Repository | B0: Data |
| AutomationML | Collada | S1: Terminology | B1: Information (3D) B6: ISM (due to kinematic definition) |
| | PLCOpen | S5: Ontology | B3: automata |
| | CAEX | S5: Ontology | B2: Information |
| STEP | | S1: Terminology | B1: Information |
| RFID | | S0: Respository | B0: Data |
| ecl@ss | | S2: Glossary | B1: Information |
| EDD | | S2: Glossary | B1: Information |
| KNX | | S3: Thesaurus | B1: Information |
| FDI | | S5: Ontology | B1: Information |
| BACNet, Enocean / BT/ Zigbee Profiles | | S3: Thesaurus | B1: Information |
| Hart / PNO / FF / SCL Profiles | | S3: Thesaurus | B1: Information |
| RDF | | S4: Ontology | B1: Information |
| RDFS | | S5: Ontology | B2: Constraints (structural) |
| CIM | | S5: Ontology | B1: Information |
| OPC UA | | S5: Ontology | B3: Automata (state machine/ filter) |
| OWL | | S5: Ontology | B2: Constraints (structural+ logical) |
| Domain Specific Language (DSL) | | S1: Terminology | B4:Automata (up to B6) |
| PackML | | S1: Terminology | B1:Information, B3:Automata |

## V. INDUSTRIE 4.0 DEMONSTRATOR FEATURES

To better understand features in the context of I4.0, we have investigated around 20 publically available demonstrators and summarized their features in Table III. Below is a condensed version of demonstrated features with their goals.

- Flexible production to increase profit in small-scale, customized production
  o Modularity on the cell-level allowing system re-layout and plug-and-play.
  o Digital product memory allowing the storage of individual production recipes and the recording product life cycle configuration.
  o Highly customized ingredients via additive printing (3D printing).
- Device-level plug-and-play to simplify engineering and commissioning efforts.
- Production monitoring via RFID or barcode, in order to increase quality of service or monitor quality of production.
- Virtual and augmented reality, in order to increase engineering and operation efficiency.
- Cross-layer and intra-layer connectivity without using traditional hierarchical structures.

By applying our generated rules in Section III, we have also listed all semantic degrees for all demonstrators in Table V. We observed that most demonstrators do not demonstrate intelligence and reasoning on the device or system level (apart from plug-and-play). Therefore, scenarios using high semantic-degrees either structurally (e.g., ontology) or behaviorally (e.g., integrated simulation model) are limited.

## VI. CASE STUDIES

To elucidate the importance of semantic information modeling for I4.0, we developed four use cases from different industry segments and transformed them into future I4.0 scenarios. These scenarios are characterized by intelligent machines or devices that support a "plug-and-produce" metaphor to enable flexible production and easy reconfiguration. Due to space limits, only concepts are demonstrated.

### A. Discrete Manufacturing

#### 1. Intelligent Devices supporting Plug-and-Play

In state-of-the-art PLC programming, a PLC program is able to access the data of the sensor via a global variable mapping process – the engineer must know how to fetch data from a sensor connected to the field bus, and later encode such information in the I/O mapping file. All these activities are tedious and error prone. Whenever an error appears in the I/O mapping process (e.g., one can accidentally misplace a pressure sensor with the adjacent temperature sensor), it is difficult to be detected. Furthermore, sensors can be configured differently (e.g., units) to support different project setups. Whenever a broken sensor is to be replaced by a new one, the new sensor needs again to be manually configured.

For the above problem, we have designed smart devices supporting plug-and-sense and experimented the system on the industrial communication protocol Modbus TCP[4]. The high-level configuration is shown in Figure 3 with entities *master*

---

[4] We use Modbus TCP, as it allows us to overlay the service network (HTTP-based) without additional hardware. However, the concept is generic and can be applied to systems running under other protocols.

*(PLCs)*, *slave (smart sensors or actuators)*, and *I4.0-service* as *logical* elements. The configuration is logical, as we can flexibly move the intelligence (e.g., logic or reasoning engine) from slaves (devices) to masters (PLCs), or to integrate I4.0-service system directly into PLCs. For ease of explanation, we explain the example of inserting a temperature sensor using the sequence diagram in Figure 4, where the temperature sensor is only able to read the temperature in Fahrenheit.

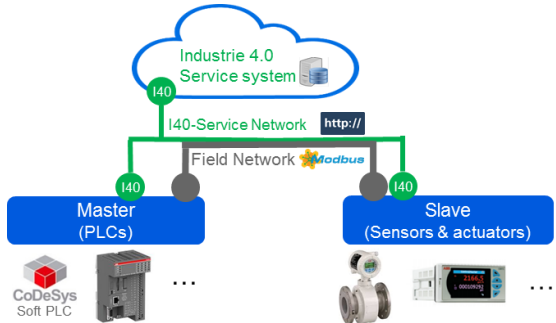

Figure 3: Intelligent devices supporting plug-and-sense under Modbus TCP communication

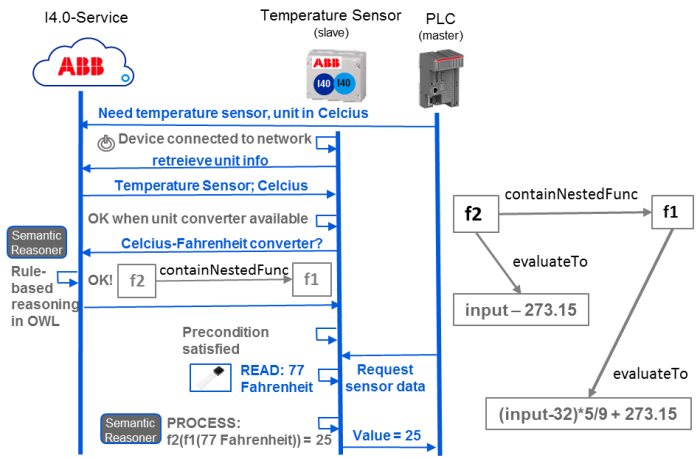

Figure 4: Sequence diagram describing the semantic reasoning when temperature sensor is inserted to the communication bus

Initially, by the time the engineer creates the I/O mapping, he also specifies what the connected device is, together with measurement units. Then the information is uploaded and stored to the I4.0-service. When a sensor is plugged to the network, it first queries the I4.0-service and examines whether itself is the desired device. In this example, if a sensor being inserted is not a temperature sensor, it immediately rejects any data requests from master. For the sensor in Figure 4, it enters a conditional acceptance mode – it is still feasible to response to the reading request when appropriate run-time converter is provided. For this purpose, the temperature sensor queries the I4.0-service in order to obtain a converter.

The I4.0-service performs a *reasoning based on transitive closures* in order to fulfill the request. This is because I4.0-service only stores the three converters: (**f1**) from Fahrenheit to Kelvin, (**f2**) from Kelvin to Celsius, and (**f3**) from Celsius to Fahrenheit. The logic reasoned establishes the knowledge that to convert from Fahrenheit to Celsius, one can first apply **f1** on the sensed value, then apply **f2** on the previously computed value. The sensor thus receives two converters **f1** and **f2**, represented by the ontological structure in Figure 4 (right). By interpreting the ontology, the sensor converts 77 Fahrenheit to 25 Celsius via functional composition **f2**(**f1**(77 Fahrenheit)), before sending it to the PLC.

| *Minimal Degree* | *Applied Rule* |
|---|---|
| S5: Ontologies | R5: The system should be dynamic and extensible by new devices and converters |
| | R7: Reasoning should be supported (the returned two functions f2 and f1 should be connected by a ontological structure such that the sensor knows how to interpret it) |
| B2: Constraints | R7: Reasoning should be supported (the sensor sends a message: "find Celcius-Fahrenheit converter", which can be viewed as a logical constraint with existential quantification) |

### 2. Deriving Machine Configurations from Production Parameters

In discrete manufacturing, individual parts are treated in multiple processing steps, typically organized in sequential production lines. For the sake of explanation, we focus on a packaging line in the food & beverage segment (Figure 5 bottom).

A Form-Fill-Seal machine fills parts produced in the upstream process into plastic bags which are packaged into cartons by a packaging machine. Finally, the cartons are placed on a pallet for shipment. Assume that there exists different product variants to be packaged which are described by different sets of production parameters. These parameters are product specific but have to be mapped onto the machine parameters of the involved machines which are standard machines provided by different vendors. Following a "plug & produce" metaphor, we expect each machine to interpret the product-specific parameters provided by a production management service and to map from product-specific to internal, machine-specific configurations. As different machines need dissimilar parameter sets, adding a new machine to the production process may require additional parameters to be handled. Missing parameters should be automatically identified and provided to a production manager for resolution.

Consider the production parameters in Figure 5 (upper part). The parameter "Weight" as defined for a product has to be interpreted in a machine specific context. The Form-Fill-Seal (FFS) machine has to interpret this parameter as the weight of the parts to be filled into the plastic bags. Hence, this parameter controls the internal dosing unit of the machine. The packaging unit talks about "Load" as the parameter to control the vacuum of the picker, which is equivalent to the parameter "Weight". The palletizer needs to understand that weight is the weight of a plastic bag and not of the packaged carton. Still the palletizer needs to know the weight of the filled carton, which it can derive from "Weight" multiplied by the "NoOfParts" (number of parts) to be packed.

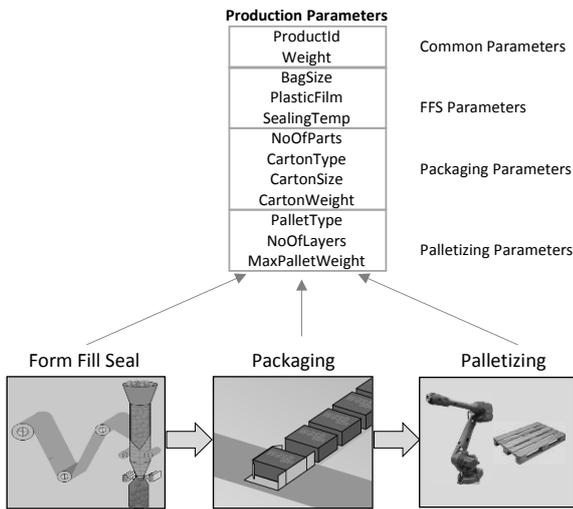

Figure 5: A sample food packaging line and corresponding production Parameters

Precisely, the reasoning from product configuration to machine configuration is first done by using semantic technologies to infer variables of different machines and extract those having the same meaning (e.g., for the above example, Packaging.Load=Product.CommonParameter.Weight). Then these inferred information is placed in the constraint system such that constraint solvers like SMT can use it to synthesize machine configurations.

| Minimal Degree | Applied Rule |
|---|---|
| S5: Ontologies | R5: The system should be dynamic and extensible by new devices and parameters. R7: Reasoning should be supported: E.g. Automatically identification of missed parameters based on process ontologies and suggestions for the missing parameters |
| B2: Constraints | R3: Is it necessary to integrate definition of terms of other parties |

### 3. Optimal Production under Interrupt

Another use case addresses exception handling in a production line. Consider the scenario where the packaging unit runs out of cartons or there is a jam in the carton supply, resulting in changing the state of the packaging unit from "Producing" to "Suspended". For optimal production, this state change is communicated to the up- and downstream machines to initiate required actions. The state "Suspended" of the packaging unit is characterized as a temporary hold which can immediately switch back to "Producing", once when the material resources have been filled up again. This is in contrast to "Aborted" which can only be resolved by forcing the packaging unit, and hence the complete line, through a reset sequence to continue production. Therefore, it is important that the up- and downstream machines have the correct semantic interpretation of the situation and adapt themselves with minimum performance degradation. The above scenario becomes even more complex if one assumes buffers appear between machines, which is considered as standardized setup for many production lines. In this case, the autonomous machines even have to consider the buffer sizes and actual usages, allowing them to gradually reduce production speed.

| Minimal Degree | Applied Rule |
|---|---|
| S1: Terminology | R1: Multiple devices the need to exchange standardized knowledge – in this case system states |
| B3: Finite automata (or B4: Petri Nets with Buffers) | R8: Is it necessary to understand/modify the functionality (e.g. logic) of another system |

### B. Process Automation

Our third example focuses on chemical plants in which products are produced in one or several processing units. Each processing unit has one or multiple input material streams where educts required for the chemical reaction are filled in. Each material supply is controlled by a charging function that is typically composed out of a pump, a valve for controlling the flow and a flow transmitter (FT) as depicted in Figure 6.

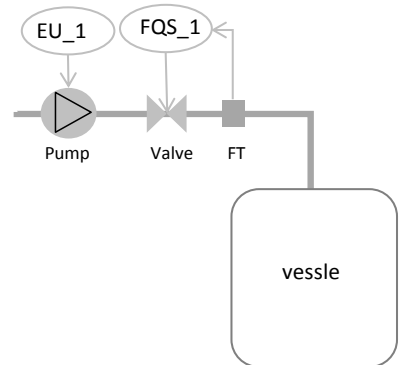

Figure 6: P&I Diagram of a Charging Unit

In I4.0, intelligent devices should autonomously organize themselves to create aggregated services based on high degree specifications such as P&I diagrams. In the diagram depicted in Figure 7, a charging function is represented by a "Flow Switch" (FQS) which stops charging the vessel when a certain amount of material has been filled. When the valve is plugged, its fieldbus signals should automatically connect to the corresponding inputs and outputs of the flow switch "FQS_1" function block. As the function block is taken from a generic application library its inputs and outputs must be mapped to the corresponding signals of the specific valve. This is where semantics can help in automatically inferring meaning of the fieldbus signals and the connectors of the function block to match both.

Similarly, when the flow transmitter is plugged, its output value should automatically connect to the process value (PV) of the function block. In this case however the units must be translated which again needs an understanding (semantics) of the physical units and rules for automatic translation. Finally, when the pump is plugged the related function block must be connected to the pump signals as well as to the flow switch function which again requires a semantic interpretation of the "Start" and "Stop" inputs of the pump.

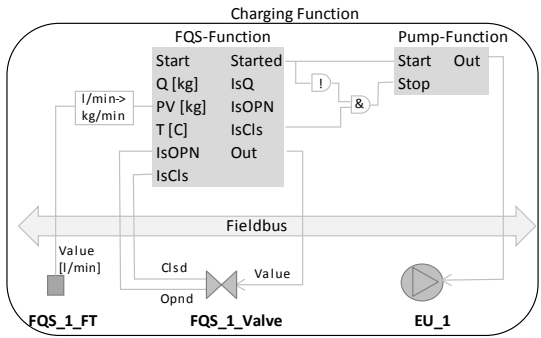

Figure 7: Functional Aggregation

| *Minimal Degree* | *Applied Rule* |
|---|---|
| S5: Ontology | R7: Reasoning on complex complex relationship is needed (not only data type – also information regarding the whole process is needed to determine if and how a device can be attached) |
| B2: Constraints (predicates over information) | R6: It is required to validate evolving configurations (e.g. of newly attached device types) during runtime. |

## VII. RELATED WORK

Using semantic technologies in the industrial context has been investigated in various subdomains. Crapo et al. studied methods to integrate semantic technologies to the power grid setup for optimization and run-time adaptation [3]. Under the context of industrial cyber-physical systems [6], Leung et al. studied the use of lattice-based ontology on ports of every component [9]. This allows engines to statically check if the dimension of the computed value is problematic due to erroneous wiring in components. There was also attempts to apply ontology within the automotive domain [11]. In discrete manufacturing, the concept of digital product memory [14] also partially includes the use of semantic technologies. Compared to individual efforts like above work, we instead offer a systematic review of categorization and the corresponding mapping of technologies. Work by Lytras and Garcia also studied how to adopt semantic technologies to industrial contexts [10], but they lack sufficient details specialized for industrial automation.

## VIII. CONCLUSION

In this paper, we identified semantic degrees and provided guidelines for engineers to select appropriate semantic degrees in their Industrie 4.0 projects. Our definition is accompanied by concrete examples in industrial contexts, enabling automation software engineers to grasp the underlying concept. Our created guidelines allow engineers to quickly identify required semantic degrees for their projects, while our evaluation over existing technologies offers a single-stop for engineer to quickly understand semantic degrees behind commonly used technologies. Our investigation over publically available demonstrators shows that most demonstrators are still largely far from intelligent, so that complex semantic degrees are not needed.

Contents of this report are already partially being used as references for creating standards for the I4.0 reference architecture. For future work, we will refine the rules and periodically update the table with new technologies and their associated semantic degrees. A decision support tool by utilizing the rules for engineering software systems in building automation is also under planning.

TABLE V. INDUSTRIE 4.0 DEMONSTRATORS AND THEIR SEMANTIC-DEGREE

| # | Demonstrator | Development | Demonstrated features related to Industrie 4.0 | Structural degree | Behavioral degree |
|---|---|---|---|---|---|
| 1 | My jogurt [15] | TU Munich TU Hamburg RWTH Aachen | 1. On every yogurt bottle, a RFID tag is attached to store the customized production recipe<br>2. Production cells automatically routes the yogurt bottle based on RFID info, without pre-planning, based on understanding the traffic of other stations<br>3. Augmented reality for remote observation of production parameters | S2: Glossary<br><br>S1: Terminology<br><br>S2: Glossary | B1: Information<br><br>B4: Petri-net<br><br>B1: Information |
| 2 | Bottle opener [16] | TU Munich & industrial partners | 1. Mobile HMI for order placement<br>2. Autonomous robotic platform carriers to transfer work pieces among work stations for customized tasks | S2: Glossary<br>S1: Terminology | B1: Information<br>B1: Information |
| 3 | Effiziente Fabrik 4.0 [17] | TU Darmstadt | 1. Network connected 3D printers for printing specialized components<br>2. Services for checking the digital product memory that shows work piece specific info such as history of production and subsequent assembly plan<br>3. Twittering sensors reporting progress and a server which aggregates relevant information | S1: Terminology<br>S2: Glossary<br><br>S2: Glossary | B1: Information<br>B1: Information<br><br>B1: Information |
| 4 | Soap factory [18] | DFKI & industrial partners | 1. Customized production via digital product memory; each station knows the sort and the amount of materials to fill in the bottle<br>2. Support reassembling on the station level | S2: Glossary<br><br>S2: Glossary | B1: Information<br><br>B2: Constraints |
| 5 | AutoPnP [19] | fortiss FESTO | 1. Flexible production via station-level re-layout<br>2. Every station provides functional interfaces (exposed as services) concerning its capabilities. When system layout changes, automatic adjust production workflow by remapping production recipe into machine instructions | S2: Glossary<br>S5: Ontology | B2: Constraints<br>B2: Constraints |
| 6 | Additive Manufacturing [20] | GE | 1. Produce metal fuel nozzles via 3D printing | S1: Terminology | B1: Information |
| 7 | F³ factory [21] | Bayer, TU Dortmund | 1. Construct small-scale modular plant via standardized process equipment assembly (PEA) such as feed (e.g., pump, storage), reaction, separation<br>2. Multiple PEAs merged into process equipment container (PEC), as one base unit for installment<br>3. PECs are designed to use standardized backbone plants such as energy, fluid pipes, process control | S1: Terminology<br><br>-<br><br>- | B1: Information<br><br>-<br><br>- |
| 8 | Key-finder production line [18] | DFKI & industrial partners | 1. Support reassembling on the station level<br>2. When system layout changes, automatic adjust production workflow by remapping production recipe into machine instructions | S2: Glossary<br>S5: Ontology | B2: Constraints<br>B2: Constraints |
| 9 | Fabrik DNA [22] | Fraunhofer IOSB | 1. Plug-and-produce over the station level; similar to AutoPnP | [see AutoPnP] | [see AutoPnp] |
| 10 | Open Integrated Factory [23] | SAP FESTO | 1. RFID on the object transportation carrier, such that products on the carrier can be customized<br>2. Seamless connectivity from ERP level to automation level<br>3. Allow querying ERP system to fetch assembly instructions<br>4. Real-time analytics for monitoring | S2: Glossary<br><br>S2: Glossary<br>S5: Ontology<br>S2: Glossary | B1: Information<br><br>B1: Information<br>B1: Information<br>B1: Information |
| 11 | Intelligente Instandhaltung [24] | SAP Harting | 1. Use Google glass for scanning and for augmented reality for remote service | S2: Glossary | B1: Information |
| 12 | Factory 2.0 [25] | GE | 1. Large scale deployments of sensors to monitor the production process and product.<br>2. Every part that goes into the batteries gets tracked with serial numbers and bar codes; allow advanced, predictive analytics | S2: Terminology<br><br>S1: Terminology | B1: Information<br><br>B1: Information |
| 13 | Injectors with IDs [26] | Bosch | 1. Injectors for trunk engine are mounted with 2D barcodes to allow production monitoring and tracking<br>2. Secure storage of product data inside corporate | S1: Terminology<br><br>S0: Repository | B1: Information<br><br>B0: Data |
| 14 | Milkrun 4.0 [27] | WITTEN-STEIN | 1. Increase productivity over logistics in a factory plant via mobile HMI<br>2. All footprints for in-factory logistics are stored and used for calculating the next optimal work piece delivery within the factory. | S2: Glossary<br>S2: Glossary | B1: Information<br>B1: Information |
| 15 | Connectivity demonstrator [28] | IBM | 1.Seamless connectivity over all layers of production from ERP (order creation) via IBM integration bus (middleware) to automation level such as PLC, via communication protocols such as MQTT<br>2. Attach RFID tags on product containers, to support flexible production | S0: Repository<br><br><br>S2: Glossary | B0: Data<br><br><br>B1: Information |
| 16 | Digital factory demonstrator [29] | Siemens HP | 1. Virtual reality plant alongside a physical production line capable demonstrating mass customization of consumer goods<br>2. Virtual commissioning - experiment flexible production, before putting into factory floor | S2: Glossary<br><br>S2: Glossary | B1: Information<br><br>B6: Simulation Model |
| 17 | Augmented Reality for Field Service | ABB | 1. Information about state of ABB devices is overlaid on a video on Android tablets at position of the devices in the room. | S2: Glossary | B1: Information |
| 18 | Smart Automation Lab [30] | RWTH Aachen | 1. Product-centric control for efficient and flexible automated one-piece flow of future automation systems.<br>2. Extensive use of CAD as virtual objects to communicate between machines | S2: Glossary<br><br>S0: Repository | B1: Information<br><br>B1: Information |